# Secure secondary utilization system of genomic data using quantum secure cloud

**Authors:** Mikio Fujiwara[1*], Hiroki Hashimoto[2], Kazuaki Doi[3], Mamiko Kujiraoka[3], Yoshimichi Tanizawa[3], Yusuke Ishida[4], Masahide Sasaki[1], and Masao Nagasaki[2*]

Correspondence: *fujwara@nict.go.jp, *nagasaki@csml.org

Affiliations:

[1]National Institute of Information and Communications Technology (NICT), 4-2-1 Nukui-kita, Koganei, Tokyo 184-8795, Japan

[2]Biosciences Unit for the Top Global Course Center for the Promotion of Interdisciplinary Education and Research (CPIER) 507, Center for Genomic Medicine, Graduate School of Medicine, Kyoto University, Kyoto 606-8507, Japan

[3]Toshiba Corporation, Corporate Research & Development Center, 1, Komukai Toshiba-cho, Saiwai-ku, Kawasaki-shi, 212-8582, Japan

[4]ZenmuTech, Inc., THE HUB Ginza OCT 804, 8-17-5 Ginza Chuo-ku, Tokyo 104-0061, Japan

**Abstract:**

Secure storage and secondary use of individual human genome data is increasingly important for genome research and personalized medicine. Currently, it is necessary to store the whole genome sequencing information (FASTQ data), which enables detections of de novo mutations and structural variations in the analysis of hereditary diseases and cancer. Furthermore, bioinformatics tools to analyze FASTQ data are frequently updated to improve the precision and recall of detected variants. However, existing secure secondary use of data, such as multi-party computation or homomorphic encryption, can handle only a limited algorithms and usually requires huge computational resources. Here, we developed a high-performance one-stop system for large-scale genome data analysis with secure secondary use of the data by the data owner and multiple users with different levels of data access control. Our quantum secure cloud system is a distributed secure genomic data analysis system (DSGD) with a "trusted server" built on a quantum secure cloud, the information-theoretically secure Tokyo QKD Network. The trusted server will be capable of deploying and running a variety of sequencing analysis hardware, such as GPUs and FPGAs, as well as CPU-based software.

We demonstrated that DSGD achieved comparable throughput with and without encryption on the trusted server Therefore, our system is ready to be installed at research institutes and hospitals that make diagnoses based on whole genome sequencing on a daily basis.

**Introduction:**

Background:

Personal human genome information is used daily in medical treatment, diagnosis of monogenic diseases, e.g., Fabry disease and Lynch syndrome [1], and prevention of side effects caused by drug administration, e.g., avoiding thiopurine treatment for inflammatory bowel disease and acute lymphocytic leukemia with specific homo alleles p.Arg139Cys in NUDT15 [2]. Additionally, personal human genome information is useful for personalized medical treatment not only for monogenic diseases but also for polygenetic diseases, e.g., type two diabetes and cardiovascular artery diseases (CAD). Statin prevention therapy for individuals with high CAD genetic risk (polygenic risk score; PRS) was estimated to reduce the mean cost per individual and improve quality-adjusted life years and averted future events of CAD [3]. The polygenic risk score is usually calculated by summing the multiplied values of the weight, i.e., the beta value from a genome-wide association study from an independent population matched CAD cohort study, and the genotype, e.g., 0, 1, or 2, in each individual. Compared to monogenic disease, to calculate PRS, genotypes from many disease-associated variants are required from every individual.

In particular, the advancement of whole-genome sequencing (WGS) technology allows us to measure the whole human genomic regions composed of 3.05 billion bases easily and on a large scale from saliva in a few days for less than $1,000. Therefore, in order to capture the high-resolution relationships between phenotypes and genotypes including rare variants, genomic cohort studies around the world are shifting from the SNP array technology, which measures the pre-designed hundreds of thousands to millions of SNPs from one sample, to the whole-genome sequencing technology. To date, more than one million samples have already been sequenced, e.g., the NIH's Trans-Omics for Precision Medicine (TOPMed) program [4], the Million Veteran Program [5], Tohoku Medical Megabank Project (TMM) [6], and UK Biobank [7]. For example, in November 2021, one large prospective cohort study, UK Biobank built a research analysis platform on the public cloud for authorized researchers to utilize WGS information of 50,000 participants [8]. In late 2023, UK Biobank plans to complete the WGS for all 500k participants [9]. The analysis of WGS data and thousands of phenotypes allow researchers to accelerate the discovery of novel variant-phenotype relationships.

The use of WGS technology for rare diseases also increases the possibility of identifying disease-causing variants. Therefore, tens of thousands individuals' whole genome sequencing project for 190 rare diseases is being conducted at Genomic England [10]. Consequently, these novel discoveries should also be frequently updated for the implementation of clinical treatment. In 2021, the American College of Medical Genetics changed the time interval for guidance on reporting secondary findings in clinical exome and genome sequencing from every four years to annually [1].

For each sample, major human whole-genome sequencing technology measures billions of short

DNA fragments as sequenced data, e.g., hundreds of DNA bases in NovaSeq 6000 (Illumina, San Diego, CA) or DNBSeq-T7 (MGI, Shenzhen, China). The human whole-genome information is composed of about 3.05 billion base pairs, and usually, the total coverage of sequenceing reads to the whole-genomic region is thirty to forty times. These billions of sequenceing reads are usually analyzed by using WGS analysis pipelines on CPUs, GPUs, or FPGAs [11]. These pipelines are frequently updated to improve the precision and recall of detected variants. In addition, the reference assembly of the human genome is updated occasionally. In 2022, T2T consortium released the new complete human genome assembly CHMv2.0 and reported improved variant detection accuracy compared to the standard reference assembly, GRCh38, through re-analysis of WGS data with CHMv2.0 [12].

Taking all of these advances and eco-system point of view into consideration, the reasonable implementation of the personal germline genome data management for medical treatment is to apply whole-genome sequencing once to a patient, store the raw WGS, i.e., FASTQ, bam, or cram, and analyze the raw WGS data using the latest reference assembly, data analysis pipeline, and the interpretation guideline. In the future, e.g., annually, if the reference assembly, data analysis pipeline, or interpretation guideline is updated, the stored raw data are reanalyzed for reinterpretation, e.g., using an additional reliable variant catalog of the patients missing in the previous pipeline, to diagnose a monogenic disease that could not be diagnosed in the previous pipeline and to improve the score of PRS.

Information Security Issues of genomic data:

A human genome is the basic information that determines an individual and cannot be modified for a lifetime. Hence protection of genomic information is required. Genome data protection techniques and various laws and guidelines are summarized in [13]. Notably, the National Institute of Health has stated that two essential values of scientific research - the need to share data broadly to maximize its use for ongoing scientific exploration and the need to protect research participants' privacy - should be balanced [14].

However, in the transmission and reception of information for genome analysis, there are many cases where information is exchanged with encryption only in the normal TLS protocol [15]. It is unlikely that its security will be compromised immediately, but it cannot be guaranteed that the information will not be decrypted decades later by a "harvest now, decrypt later attack" with to the improved performance of quantum computers and supercomputers in the future [16]. At the transmission and storage of data that requires long-term confidentiality, such as genomic data, it is required to have sufficient security against future computers and cryptanalysis algorithms. In recent years, laws that impose severe civil punishment for leakage of personal information have been enforced [13], and it is desired to build a secure system that can absolutely eliminate the threat of future decryption in line with the growing awareness of personal information protection (in short, an information theoretically secure system).

Solution for genomic data protection:

As a solution to this need, quantum key distribution (QKD) [17,18], which allows sharing information using theoretically secure keys between two parties, has been attracting attention in recent years. Vernam's one time pad (OTP) [19] encryption using the key from a QKD link enables information theoretically secure data transmission and eliminates concerns about future decryption. On the other hand, for example, in the QKD using the BB84 protocol, a single photon is used as the key transmission medium, so it is easily affected by transmission loss, and the transmission distance is limited to about 50 to 100 km in the field. However, current QKD networks enable to expand the key supply length by performing key relay via a "trusted node" [20]. Moreover, a distributed storage system has been built on the QKD network and realized information theoretically secure data storage [20]. From the viewpoint of information security, such a distributed storage system on the QKD network would be suitable for storing data that requires confidentiality for a very long period, such as personal genomic data. This secure distributed storage system has recently been updated with an enhanced computing functionality for secure secondary use of data, referred to as a "quantum secure cloud". According to secure secondary use of data, several solutions have been proposed, e.g., secure computation using multi-party computation [21,22] and homomorphic encryption [23]. Unfortunately, each of these two schemes has its own issues, such as increased communication volume and computational complexity. These issues result in a degradation of throughput.

Preset work for secure secondary use of data methods (multi-party computation and homomorphic encryption)

Multi-party computation requires a huge amount of computational and communication resources. Computational resources increase by the number of shares [21,22], and communication between the data owner and share holders would be a rate-determining process.

Research aiming for both communication efficiency and confidentiality in multiparty computation schemes is underway. Zhao et al., [22] described a survey of recent activities and introduced implementation schemes targeting several genomic data sets. In [24], three secure sequence comparison protocols were proposed based on garbled circuit techniques and under a semi-honest model. However, inefficiencies of their schemes were mentioned when dealing with large amounts of data. A theoretical study of efficiency improvement using secure two-party computation (S2PC) was reported in [25], Although more efficient than conventional systems, it is not sufficient from the standpoint of practicality. BesidesS2PC implementations for genome data comparison were proposed in [26,27], their methods are different from our method for whole genome data analysis. It is also not consistent with our policy of information theoretic security, which we consider essential for genomic data analysis.

As a homomorphic encryption scheme, the throughput is not sufficient because the calculation is complicated [28]. As a scheme to mitigate the computational complexity of homomorphic encryption, a method called switchable homomorphic encryption (SHE), which can combine both additive and multiplicative schemes when needed, was proposed [29]. And in [30], Genome-wide association studies (GWASs) analysis using homomorphic encryption was demonstrated. Moreover, FPGA and ASIC hardware solutions would boost the throughput of homomorphic encryption, while such efforts are still in the development stage [31,32]. These implementation innovations have the potential to expand the use of homomorphic encryption. However, throughput improvements are in the process of development and information theoretic security has not been achieved.

In addition to high throughput technology, since a large amount of data is used in genome analysis, e.g., whole-genome sequencing data from one individual is usually more than 30 GB bases in FASTQ format, both multi-party computation and homomorphic encryption need to save computational and communication resources. Secondly, genomic data, e.g., a FASTQ file, is categorized to unstructured data [33] since it is not suitable for secure computation by multi-party computation and homomorphic encryption.

Our approach; using a trusted server for secure secondary use of data

To improve throughput, secure computation using a "trusted server" enables realistic and secure data use, and has been proposed as one form of implementation [34]. Essentially, it is necessary to have a one-stop system that can guarantee information theoretic security among the data owner, i.e., the authorized genomic data bank, and users, e.g., the medical doctor.
Towards the implementation of a one-stop system, for the first step, we developed a method to guarantee the integrity of distributed storage data with information theoretical security by assuming a "trusted share calculator" and a verifier in the quantum secure cloud [35]. In this paper, based on this "trusted server" in the quantum secure cloud concept, we developed an advanced high-performance one-stop system for large-scale genome data analysis with secure secondary use of data to the data owner and multiple users with different levels of data access control.

This manuscript is organized as follows. In Materials and Method, we introduce our experimental setup i.e. the quantum secure cloud and the implementation of the trusted server. In Results, the throughputs of our system are reported. We summarize our experiment and discuss the outlook of our system in Conclusion.

**Materials and Methods**

Quantum Secure Cloud System

   Our quantum secure cloud system is a distributed secure genomic data analysis system (DSGD) built on the Tokyo QKD Network. The Tokyo QKD Network [36] is established on the National Institute of Information and Communications Technology (NICT) optical fiber testbed. The QKD network works as a secure key supply infrastructure. Figure 1 (b) shows a conceptual diagram of the Tokyo QKD Network by referring to the open system interconnection (OSI) reference model. The Tokyo QKD Network has five trusted nodes (cylinders in figure), and each node is connected by QKD links (blue lines) in the quantum layer (bottom layer in figure) and key management systems with public channels (green lines) in the key management layer (middle layer in figure). The five trusted nodes are physically distributed about 100 km between NICT (nodes 1 to 4) and Otemachi (node 5) by various vendors [37-41]. Each node works as a "trusted node" and the keys from the QKD links are strictly managed. Each trusted node has a key management agent (KMA) and a key supply agent (KSA) (box in Figure 1) and located in a physically protected place. Keys generated in each QKD link in the quantum layer are pushed up (dashed line in Figure 1) to the KMA in the key management layer and transferred to the KSA. In the key management layer, the key management server (KMS) gathers link information and instructs KMAs to execute key relay according to requests from the DSGD.

  As the service layer, the DSGD realizes a high-performance one-stop system for large-scale genome data analysis with secure secondary use of data to the data owner of genomic data and multiple users with different levels of data access control. The data owner has the administrative privilege of depositing genomic data and controlling secondary use of the data on the DSGD. As in Figure 1 (a), the DSGD stores genomic data, e.g., FASTQ, cram, bam, VCF files, on a distributed storage system (the triangle system on the left), transfers and decodes the encrypted data to a trusted server (top of the figure), and reports the filtered (part of) genotype information, e.g., the minimum genotypes of the patient for diagnosis, to authorized users (right side of the figure). For the distributed data encryption and decryption in the service layer, OTP encrypted communication is used between the key management layer and the service layer via KSAs.

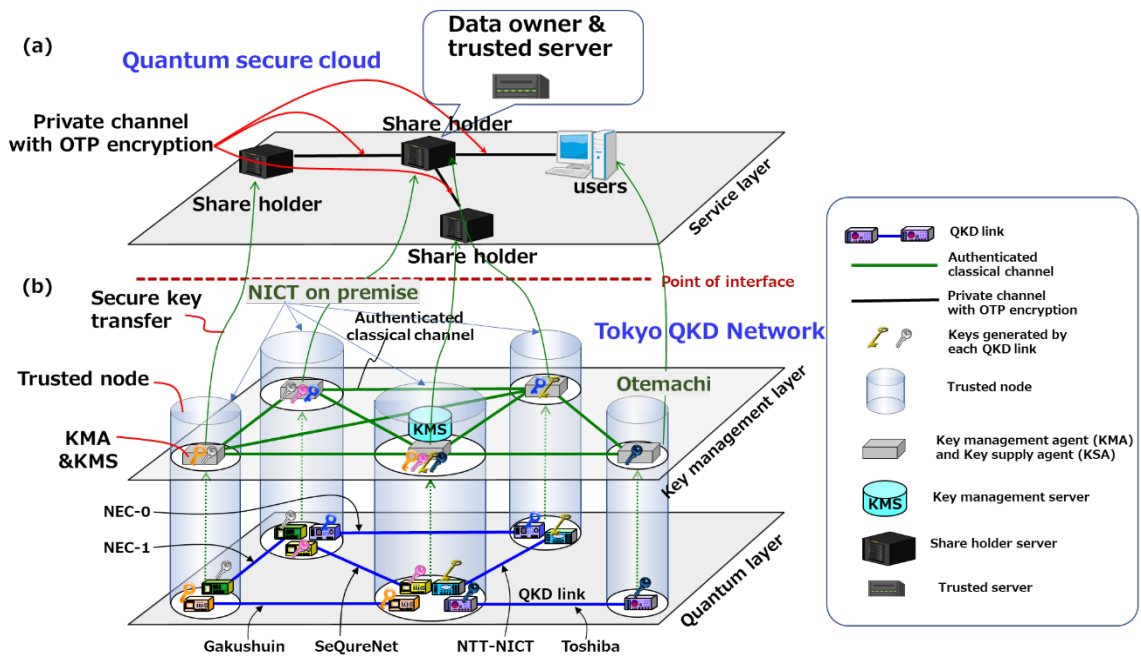

**Fig 1. Conceptual view of Tokyo QKD Network.**

The QKD network works as a secure key supply infrastructure. (a) Service layer; Secret sharing or other services are installed in this QKD network. The data owner, the trusted server, share holders, and the user at the service layer communicate through OTP encrypted communication lines in which secure keys are provided from the QKD network. Once supplied with the keys or random numbers, the key data in the QKD network are erased and the responsibility of key management moves to users at the service layer. (b) quantum layer and key management layer; Generated keys in each QKD link are pushed up to servers, called key management agents (KMAs). Each KMA is set in a physically protected place, referred to as a "trusted node". A key supply agent (KSA) is integrated to the KMA. The KSAs supply users with the keys. A key management server (KMS) gathers link information and instructs KMAs to execute key relay according to request from the service layer.

Distributed System in Service Layer

We have implemented the secret sharing protocol of Shamir's scheme [42] in the service layerand realized information theoretically secure data transmission, storage, authentication, and restoration [21]. This time, to deal with a large amount of genomic data, we implemented an XOR-based secret sharing that is simpler in calculation than Shamir's scheme while having information theoretic security and is expected to realize high throughput. The method we implemented is the scheme with the (2,3) threshold.

For the secret data $S = S_1 \cdot S_2$ ($\cdot$ in combination, the data size of each data is the same), we prepare random numbers $R = R_1 \cdot R_2$ with the same number of bits, and calculate three shares as follows; $A = A_1 \cdot A_2 = (S_1 \oplus R_1) \cdot (S_2 \oplus R_2 \oplus R_1)$, $B = B_1 \cdot B_2 = (S_1 \oplus R_1 \oplus R_2) \cdot (S_2 \oplus R_2)$, $C = C_1 \cdot C_2 = R_1 \cdot R_2$.

XOR-based secret sharing can achieve high throughput with information theoretic security, because the calculation of XOR is simple. But it does not have the additive homomorphic nature of shares like Shamir's scheme, and secure computation based on multi-party computation is impossible. There is an example of high-speed secure computation of multi-party computation [43]. This scheme is based on arithmetic calculation with a (2,3) threshold secret sharing scheme and XOR is used in a particular case on the ring of modulo 2. However, the size of the share is twice the original data (6 times the total data), and it will be necessary to consider the cost of communication and computation when secret sharing large volumes of data. The secret sharing based on XOR would be frequently used for distributed processing of a large amount of data because it does not require the development of special software and is superior in terms of simplicity.

Figure 2 shows the secret sharing and secure computation configuration implemented on the Tokyo QKD Network.

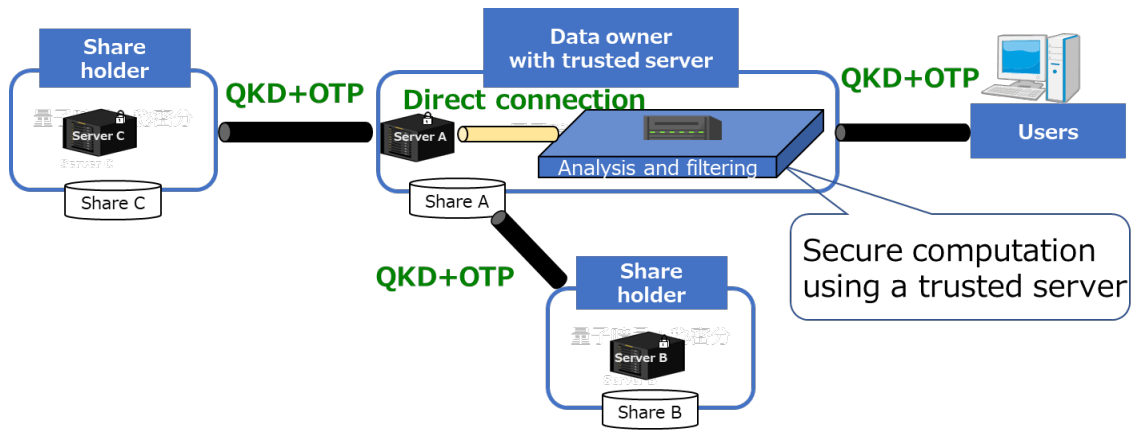

**Fig. 2. The network configuration of secret sharing and secure computation on the Tokyo QKD Network.**

The share holders (Share A, B and C) and the data owner are established in trusted nodes of Tokyo QKD Network. The data owner has a function of the share holder. The trusted server is set in the same place of the data owner. The share holders, the data owner, and the users communicate by using OTP encryption from keys generated in the Tokyo QKD Network.

Definition of Trusted Server in Service Layer

Since large volume data with a block structure such as FASTQ should be treated in the same way as "unstructured data" due to its large volume, so it is extremely difficult to achieve a user-satisfactory throughput by multi-party computation or homomorphic encryption. In other words, we need a system that allows efficient and secure secondary use genomic data. Given these conditions, we propose to establish a trusted server in the quantum secure cloud to enable the secondary use of secure data distributed by the XOR-based secret sharing.

QKD networks around the world assume "trusted nodes" and achieve key relays to expand the service area. Since such QKD networks are established on physical protection nodes, the secondary use of data assuming "a trusted server" in the quantum secure cloud has a very high affinity conceptually.

Ideally, similar requirements of hardware security module (HSM) defined in FIPS140-2 would be required as the conditions to be a trusted server. With reference to the HSM specification requirements, the following conditions are implemented in our trusted server in Table 1.

**Table1. Conditions of the trusted server in our experiment.** Similar requirements of hardware security module (HSM) defined in FIPS140-2 would be required as the conditions to be "a trusted server." Refer to the HSM requirements, conditions listed in Table 1 were implemented in this demonstration.

| 1 | Access rights to the server are strictly managed |
| --- | --- |
| 2 | In the case of external connection, hardware authentication must be carried out in an information theoretically secure manner. |
| 3 | A protected area of a certain capacity (e.g. 1 GB) can be assumed in the server. |
| 4 | Do not store unencrypted data in the protected area of the server for long periods (e.g. 24 hours or more). |
| 5 | Must be installed in a server room with strict access control. |
| 6 | Unnecessary external interface is disabled. |
| 7 | If you want to store data outside for a long time, carry out secret sharing. |
| 8 | OTP encryption when transmitting data to the outside. |
| 9 | Make sure to erase the random numbers used for encryption and secret sharing. |
| 10 | Use a random number source that does not have periodicity, such as a physical / quantum random number generator. |
| 11 | Other security on general information systems such as intrusion detection system (IDS) and intrusion prevention system (IPS) should be applied. |

Implementation of Trusted Server in Service Layer

The above conditions are implemented in a dedicated server for genome data analysis to realize secure secondary use of data. We show our secure secondary use system installed in the Tokyo QKD Network.

Figure 3 shows the diagram of a trusted server. We used DRAGEN [44] as a computational engine in the server for genome analysis. DRAGEN enables the generation of a VCF file, which is a set of conventions for representing all sites within the genome in a reasonably compact format from a FASTQ file, which is a text file that contains the sequence data from the clusters that pass a quality filter on a flow cell effectively. When a data request comes from an end user, the data owner who has share A asks one of the share holders to send back the share (B or C), and re-constructs the data (FASTQ file). When the devices are connected between the data owner and the end user, Wegman-Carter authentication [45] is carried out, which is information theoretically secure. The FASTQ file is input to DRAGEN and is transformed into a VCF file. When the data owner sends the VCF file by OTP, the data owner controls the disclosure part of the VCF file according to the access right of the end user. The VCF files include all sites within the region of interest in a single file for each sample. There is a risk of identifying an individual from mutation information for non-research purposes contained in the VCF file. Therefore, to prevent unnecessary leakage of personal information, the data owner controls the range of mutants to be disclosed according to the user's access rights. In the next section, we describe the performance of our system.

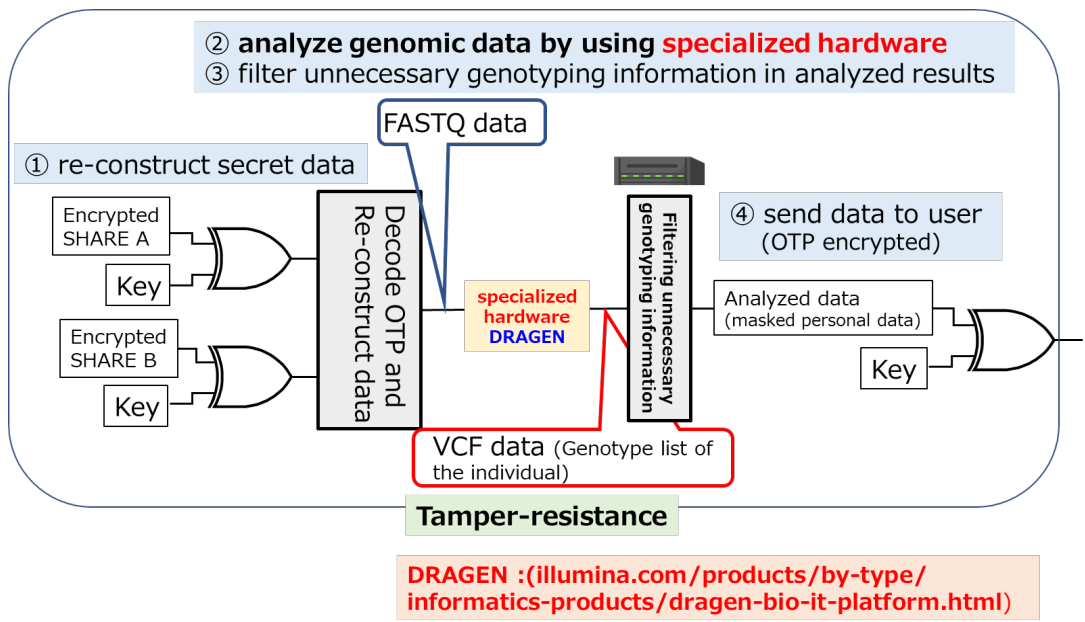

**Fig. 3. Diagram of the trusted server for genomic data analysis**

DRAGEN is used as a trusted server for genome analysis in our reference implementation. After reconstructing encrypted data as FASTQ using shares, the trusted server takes the FASTQ and generates variant call format (VCF) file, which is a list of genotypes different from the reference human genome assembly, e.g., three to four million genotyping records from an individual FASTQ file. The part of VCF file, is sent to the user after filtering the unnecessary genotyping information. This trusted server is installed in a physically protected area with tamper-resistance.

## Results

### Benchmarking of DSGD

The secret sharing and secure computation were established on the Tokyo QKD Network, and the throughput was measured including the genotype filtering function. A FASTQ file has a volume at the 10 GB level, but it is compressed to about 400 MB when converted to a VCF file. Furthermore, it is operationally assumed that the disclosure range changes from 50 MB to less than 1 MB by filtering. The definition of throughput was measured according the following five equations;

$$A = \frac{size(FASTQ) + size(VCF)}{processing\ time} \quad (1)$$

$$B\text{-}1 = \frac{size(VCF.gz) + size(VCF)}{processing\ time} \quad (2)$$

$$B\text{-}2 = \frac{size(VCF) + size(F\_VCF)}{processing\ time} \quad (3)$$

$$B\text{-}3 = \frac{size(F\_VCF) + size(F\_VCF.gz)}{processing\ time} \quad (4)$$

$$\text{Total throughput} = \frac{size(FASTQ)}{total\ processing\ time} \quad (5)$$

Processing time include data format transform.

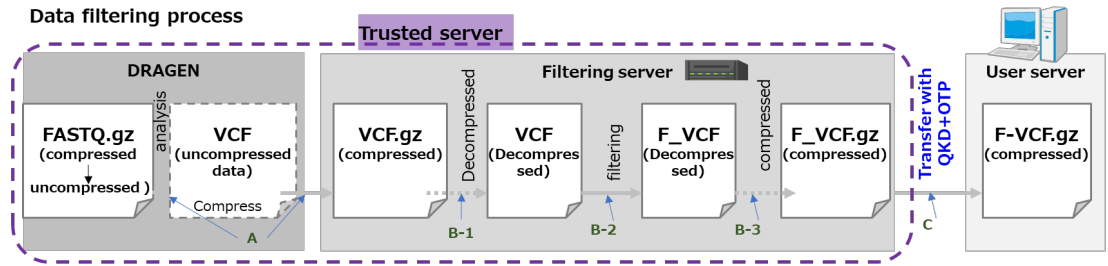

**Fig. 4. Definition of throughput of secure computation using the trusted server. Its processing**

**time includes data format transform.**

The trusted server consists of DRAGEN and a filtering server. The DRAGEN takes a decrypted FASTQ.gz file as input, processes the file to generate a VCF file, then generates a GZIP compressed VCF file (VCF.gz) (throughput A). The VCF.gz file is transferred to the filtering server. On the filtering server, the requested genotype regions of the VCF.gz file is decompressed (throughput B-1), filtered (throughput B-2), compressed (throughput B-3), and transferred to the server of an authenticated user (throughput C). The equations of throughputs A, B-1, B-2, B-3, and C, i.e., X[Mbit], are defined at the bottom of this figure.

Assuming cases (extracting 0, 30,000, 1.5 million, and 3 million cases) of filtering ranges, the throughput in each case was measured. The results are summarized in Table 2. Table 3 focuses on the throughput of OTP transmission (encryption and decryption processes).

**Table 2. Throughput in each process. Data storage type in each process is "share", the process to restore the original data is also included. Transfer condition plain: No encryption OTP: Vernam's one time pad Encryption and Decryption**

Data storage type in each process is "share", the process to restore the original data is also included.

Plain: No encryption, OTP: Vernam's one time pad Encryption/Decryption, respectively.

"Normal" means data itself. A and B correspond to A and B in Fig. 4.

| processing | | Data storage type | Transfer condition (among share holders) | Data volume [MB] | OS RHEL8.4 Processing time [s] | throughput [Mbps] |
|---|---|---|---|---|---|---|
| A | | normal | plain | 23,498 | 340.2 | 536 |
| | | share | plain | 23,498 | 352.1 | 518 |
| | | share | OTP | 23,498 | 355.5 | 513 |
| B | 1 | normal | plain | 935 | 3 | 2,474 |
| | | share | plain | 935 | 3.8 | 1,948 |
| | | share | OTP | 935 | 3.2 | 2,318 |
| | 2 | normal | plain | 1,232 | 9.6 | 976 |
| | | share | plain | 1,232 | 9.8 | 954 |
| | | share | OTP | 1,232 | 9.9 | 940 |
| | 3 | normal | plain | 466 | 7.2 | 436 |
| | | share | plain | 466 | 7.4 | 425 |
| | | share | OTP | 466 | 7.3 | 427 |
| VCF (extracted sample: 1.5 million) | | normal | plain | 56 | 0.9 | 411 |

|  |  |  |  |  |  |
|---|---|---|---|---|---|
|  | share | plain | 56 | 0.9 | 414 |
|  | share | OTP | 56 | 1.3 | 285 |
| Total throughput (extracted sample: 0) | normal | plain | 22,676 | 352.5 | 515 |
|  | share | plain | 22,676 | 365.6 | 496 |
|  | share | OTP | 22,676 | 368.6 | 492 |
| Total throughput (extracted sample: 30 thousand) | normal | plain | 22,676 | 352.8 | 514 |
|  | share | plain | 22,676 | 365.9 | 496 |
|  | share | OTP | 22,676 | 368.9 | 492 |
| Total throughput (extracted sample: 1.5 million) | normal | plain | 22,676 | 360.9 | 503 |
|  | share | plain | 22,676 | 374 | 485 |
|  | share | OTP | 22,676 | 377.2 | 481 |
| Total throughput (extracted sample: 3.0 million) | normal | plain | 22,676 | 371.6 | 488 |
|  | share | plain | 22,676 | 385.3 | 471 |
|  | share | OTP | 22,676 | 389.4 | 466 |

**Table 3. Throughput of OTP transmission**

The throughput with and without OTP encryption is summarized. The communication protocols with TCP and UDP are listed. "Header encryption" means OTP encryption including header information of each protocol. Maximum transmission unit corresponds to the data size of IPSEC.

|  |  | OTP |  | Plain |  |
|---|---|---|---|---|---|
| Header encryption |  | NO | YES | NO | YES |
| Maximum transmission unit (BT) |  | 1470 | 1,454 | 1,470 | 1,454 |
| TCP(Gbits/sec) |  |  |  |  |  |
|  | 1 | 2.87 | 1.61 | 2.73 | 2.74 |
|  | 2 | 2.89 | 1.62 | 2.66 | 2.89 |
|  | 3 | 2.91 | 1.61 | 2.85 | 2.85 |
|          | 4 | 2.91 | 1.59 | 2.63 | 2.82 |
|          | 5 | 2.86 | 1.62 | 2.67 | 2.83 |
| average  |   | 2.89 | 1.61 | 2.71 | 2.83 |
| UDP (Gbits/sec) | | | | | |
|          | 1 | 3.34 | 3.24 | 3.49 | 3.41 |
|          | 2 | 3.33 | 3.23 | 3.45 | 3.41 |
|          | 3 | 3.32 | 3.25 | 3.43 | 3.39 |
|          | 4 | 3.35 | 3.28 | 3.44 | 3.28 |
|          | 5 | 3.41 | 3.31 | 3.49 | 3.46 |
| average  |   | 3.35 | 3.26 | 3.46 | 3.39 |

A throughput exceeding 400 Mbps was achieved by secret sharing on the Tokyo QKD Network and secure computation using the trusted server. These results mean that the throughput is limited by the processing time in DRAGEN. Our system can provide genomic analysis data to users without the additional latency associated with data concealment. To our knowledge, the operation cannot be handled on multi-party computation or homomorphic encryption. In addition, the filtering function in this system enables us to conduct genome analysis research without worrying about the leakage of unnecessary personal data.

Security Enhancement of Trusted Node

We have also made various efforts to implement security of the trusted server. In particular, erasing the used key is a necessary function for OTP encryption, while careful consideration is required to realize it. For example, solid state devices (SSDs), which are widely used as data storage media, can be restored in many cases even if the data is erased at first glance, suggesting that there is a risk of data leakage [46]. In that respect, the data of random access memories (DRAMs) is surely erased when the power is turned off. If all the key data for OTP can be stored in DRAMs, it is certain in terms of erasing the data. However, there is a risk that valuable keys from QKD links will be lost due to instantaneous electric outages, etc., and there is anxiety about stable operation. To solve these problems, the method shown in Fig. 5 is also used to erase random numbers used at OTP transmission in our system [47].

The detailed procedure is as follows. The key K0 provided by the QKD link is separated into K1 and K2 (size (K1) >> size (k2)). K1 is stored in an SSD or another medium that can be stored for a long time, while K2 is stored in a DRAM. Key expansion using the AES scheme is performed using K2 as the initial random number, and the key is stored in DRAM as K3. At that time, set size (K1) = size (K3). Furthermore, the XOR of K1 and K3 is calculated as K4, and K4 is also stored in the DRAM. K4 is the key for OTP transmission. By using this method, it has information theoretic security against eavesdropping on the communication path. And even if the SSD that should have been erased is stolen by a malicious third party, it is extremely difficult to guess K4. This method is computationally secure against SSD attacks, however, implementation becomes much safer.

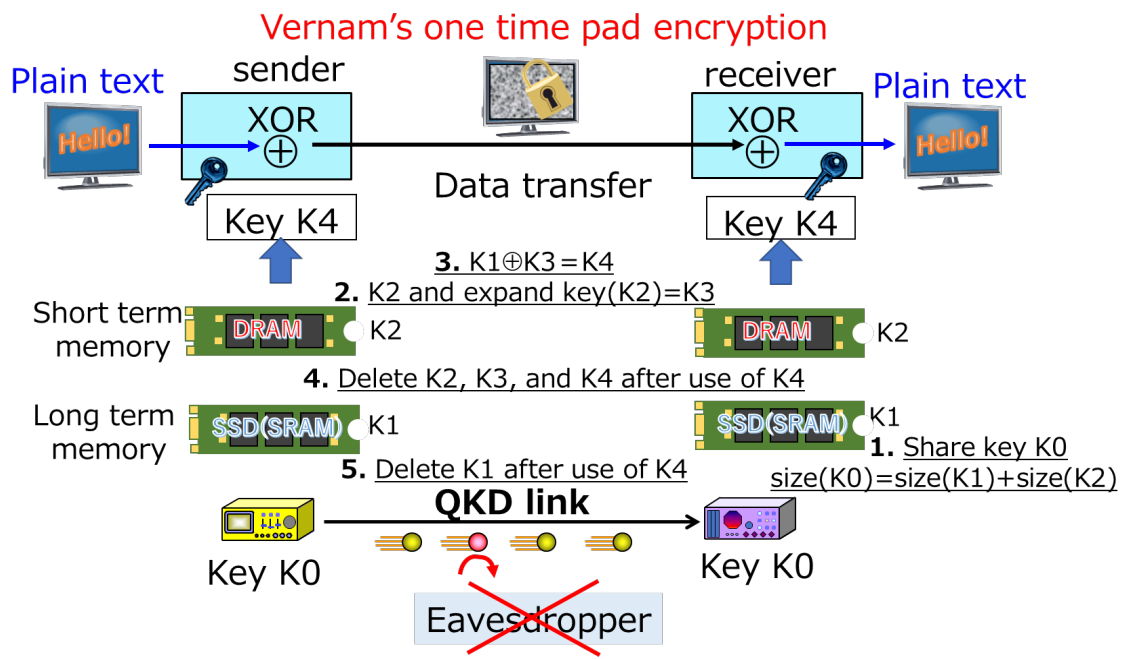

**Fig. 5. Key generation and erasing method for OTP encryption**

Steps 1 to 3 are the key generation steps and Steps 4 and 5 are the key erasing steps for OTP encryption.

Step 1: the key K0 provided from the QKD link is separated into K1 and K2 (size (K1) >> size (K2)). Step 2: K1 is stored in an SSD (SRAM) or other medium that can be stored for a long time, while K2 is stored in a DRAM (short term memory). Key expansion using the AES scheme is performed using K2 as the initial random number, and the key is stored in DRAM as K3. At that time, set size (K1) = size (K3) on DRAM. Step 3: the XOR of K1 and K3 is calculated as K4. Step 4: Delete keys in DRAM Step 5: Delete keys in SSD. By completing Steps 4 and 5, impossible to guess the K4 by the probing attack to the SSD.

A fuzzing test was conducted on the network around the trusted server, and measures such as general cyber countermeasures were implemented. In addition, the rack that mounts the trusted server has a lock function by face recognition, and our experimental system realized a defense-in-depth implementation. It is difficult to make a general conclusion on how secure our experimental implementation is because we need to confirm the existence of security holes in other network devices. However, we can conclude that we succeeded in proof-of-principle demonstration of secure computation using a trusted server with information theoretical security.

Genome data i.e. FASTQ file is structured to some extent, but it does not apply to statistical processing like the case of electronic medical record data. Therefore, analysis by a dedicated analysis device is adopted worldwide. We think that it is difficult to apply multi-party computation or homomorphic encryption to such genomic data whose volume is large and is not well structured. Therefore, we think that the secure computation method using a trusted server in a secure network capable of information theoretically secure transmission and storage is a practical and promising candidate for secure secondary use of genomic data.

**Conclusion**

As a method to enable information theoretically secure genomic data transmission, storage, and secondary use, we proposed and implemented a secret sharing system on a quantum key distribution network and secure secondary use method of data assuming dedicated hardware as a trusted server. We have realized high-speed processing of over 400 Mbps for data analysis to obtain genotyping information from GB-class sequencing genomic data and made it possible to hand over to authorized users in an information theoretically secure manner using OTP transmission and secret sharing. From the perspective of enabling the secure secondary use of genomic data, the multi-party computation would be a candidate, however, it cannot always be implemented with a realistic throughput. In addition, while multi-party computation requires as many computational resources as there are shares, our method is equivalent to ordinary computation. In addition, since the analysis method for genome analysis is updated frequently, it is necessary to update the software with the same number of computation resources as the same of shares. On the other hand, when we use the scheme of assuming a trusted server, it can be completed by updating the software and firmware of one dedicated hardware, so the maintenance work can be minimized, and there are many advantages in terms of cost and operation. Research and diagnosis using human genome data must maintain secrecy for a very long time. The combination of QKD network, secret sharing, and secondary use of data by a trusted server is extremely effective as a practical implementation method for this purpose. To make this scheme more mature, it is expected that security certification methods will be required. The penalty for leakage of personal information is increasing, which leads to the high cost of using personal information. If the personal information is aggregated in the quantum secure cloud and the secure computation scheme proposed in this study is used, it is possible to reduce the cost of protecting personal information at hospitals or institutes. In the future, we would like to complete a system that allows various medical institutions to store data with information theoretic security and allows cross-referencing and highly efficient secondary use of data by multiple organizations using our proposed system.

**References:**

[1] Miller, D. T., "ACMG SF v3.0 list for reporting of secondary findings in clinical exome and genome sequencing: a policy statement of the American College of Medical Genetics and Genomics (ACMG)," Genet Med, 23, 1381-1390, (2021).

[2] Kakuta, Y, et al., "NUDT15 codon 139 is the best pharmacogenetic marker for predicting thiopurine-induced severe adverse events in Japanese patients with inflammatory bowel disease: a multicenter study," J Gastroenterol, 53, 1065-1078, (2018).

[3] Mujwara, D., et al., "Integrating a Polygenic Risk Score for Coronary Artery Disease as a Risk-Enhancing Factor in the Pooled Cohort Equation: A Cost-Effectiveness Analysis Study," J Am Heart Assoc, (2022). Online ahead of print.

[4] Taliun, D., et al., "Sequencing of 53,831 diverse genomes from the NHLBI TOPMed Program," Nature, 590, 290-299, (2021).

[5] Gazianoab, J. M., et al., "Million Veteran Program: A mega-biobank to study genetic influences on health and disease," J Clin Epidemiol, 70, 214-223, 2016.

[6] Kuriyama, S., et al., "The Tohoku Medical Megabank Project: Design and Mission," J Epidemiol, 26, 493-511, (2016).

[7] Sudlow, C., et al., "UK biobank: an open access resource for identifying the causes of a wide range of complex diseases of middle and old age," PLoS Med, 12, e1001779, (2015).

[8] https://allofus.nih.gov/news-events/announcements/program-releases-first-genomic-dataset Date of access: 06/24/2022.

[9] https://www.ukbiobank.ac.uk/enable-your-research/about-our-data/future-data-release-timelines Date of access: 06/24/2022.

[10] Smedley, D., et al., "100,000 genomes pilot on rare-disease diagnosis in health care - preliminary report," N Engl J Med, 385, 1868-1880, (2021).

[11] Tanjo,T., Kawai, Y., Tokunaga, K., Ogasawara, O., and Nagasaki, M., "Practical guide for managing large-scale human genome data in research," J Hum Genet, 66, 39-52, (2021).

[12] Nurk, S., et al., "The complete sequence of a human genome," Science, vol. 376, pp.44-53, 2022.

[13] Wang, Z., Hazel, J. W., Clayton, E. W., Corobychik, Y., Kantarcioglu, M., and Malin, B. A., "Sociotechnical safeguards for genomic data privacy," Nature Reviews Genetics, https://www.nature.com/articles/s41576-022-00455-y (2022).

[14] https://www.genome.gov/about-genomics/policy-issues/Privacy Date of access: 06/24/2022.

[15] private communication with Prof. Miyano (2018).

[16] e.g. https://www.freedomlab.com/posts/harvest-now-decrypt-later Date of access: 07/15/2022.

[17] Bennett, C. H. & Brassard, G. Quantum cryptography: Public-key distribution and coin tossing,

Proceedings IEEE Int. Conf. on Computers, Systems and Signal Processing, Bangalore, India, pp. 175–179 (IEEE, New York, 1984).

[18] Gisin, N., Ribordy, G., Tittel, W. & Zbinden, H. Quantum cryptography. Rev. Mod. Phys., 74, 145–195 (2002).

[19] Vernam, G. S., "Cipher printing telegraph systems for secret wire and radio telegraphic communications," J. American Institute of Electrical Engineers, 45, 295-301, (1926).

[20] ITU-T Y.3800 (10/2019).

[21] Fujiwara, M., Waseda, A., Nojima, R., Moriai, S., Ogata, W., and Sasaki, M., "Unbreakable distributed storage with the quantum key distribution network and password-authenticated secret sharing," Sci. Reports, 6, 28988-1-8, (2016).

[22] Zhao, C., Zhao, S., Zhao, M., Chen, Z., Gao, C.-Z., Li, H., and Tan, Y., "Secure multi-party computation: Theory, practice and applications," Information Science 476, 357-372 (2019)

[23] .Cho, H., Wu, D. J. & Berger, B. Secure genome-wide association analysis using multiparty computation. Nat. Biotechnol. 36, 547–551 (2018).

[24] Jha, S., Kruger, L., and Shmatikov, V., "Towards practical privacy for genomic computation," Proceedings of the IEEE Symposium on Security and Privacy 216–230 (2008).

[25] Huang, Y., Evans, D., Katz, J., and Malla, L., "Faster secure two-party computation using garbled circuits," Proceedings of the USENIX Security Symposium, 201, 331-335 (2011).

[26] Wang, S., X., Huang, Y., Zhao, Y., Tang., H., Wang., X., and Bu, D., "Efficient genomic-wide, privacy-preserving similar patient query based on private edit distance," Proceedings of the 22nd ACM SIGSAC Conference on Computer and Communications Security, 492–503 (2015).

[27] Zhu, R., and Huang, Y., "Efficient privacy-preserving general edit distance and beyond," https://eprint.iacr.org/2017/683.pdf.

[28] Ayday, E., Raisaro, J. L., Hubaux, J.-P. & Rougemont, J. Protecting and evaluating genomic privacy in medical tests and personalized medicine. Proc. 12th ACM Workshop Priv. Electron. Soc. 2013, 95–106 (2013).

[29] Lim. H., W., Tople, S., Saxena, P., and Chang, E.-C., "Faster secure arithmetic computation using switchable homomorphic encryption," https://eprint.iacr.org/2014/539.pdf.

[30] Blatt, M., Gusev, A., Polyakov, Y., and Goldwasser, S., "Secure large-scale genome-wide association studies using homomorphic encryption," PNAS, 177(21), 11608-11613 (2020).

[31] Chillotti, I., Gama, N., and Izabachene, M., "Improving TFHE: faster packed homomorphic operations and efficient circuit bootstrapping," https://eprint.iacr.org/2017/430.pdf.

[32] Reis, D., Takeshita, J., Jung, T., Niemier, M., and Hu, X. S., "Computing-in-memory for performance and energy-efficient homomorphic encryption," IEEE Trans. VLSI system.vol.(28), no.(11), 2300-2313 (2020).

[33] e.g. https://www.techtarget.com/searchbusinessanalytics/definition/unstructured-data

Date of access: 07/15/2022.

[34] Bomhard, N. von, Ahlborn, B., Masson, C., and Mansmann, U., "The trusted server; a secure computational environment for privacy compliant evaluations on plain personal data," PLOS ONE, 0202752, September 6, 1-19, (2018).

[35] Fujiwara, M., Nojima, R., Tsurumaru, T., Moriai, S., and Sasaki, M., "Long-Term Secure Distributed Storage Using Quantum Key Distribution Network With Third-Party Verification," IEEE trans. Quantum Engineering, vol.3, No. 4100111, (2022). DOI: 10.1109/TQE.2021.3135077

[36] Sasaki, M., et al., "Field test of quantum key distribution in the Tokyo QKD Network," Opt. Express 19(11), 10387–10409, (2011).

[37] Yoshino, K., Ochi, T., Fujiwara, M., Sasaki, M. & Tajima, A. Maintenance-free operation of WDM quantum key distribution system through a field fiber over 30 days. Opt. Express 21, 31395-31401 (2013).

[38] Dynes, J. F. et al. Stability of high bit rate quantum key distribution on installed fiber. Opt. Express 20, 16339-16347 (2012).

[39] Shimizu, K. et al. Performance of long-distance quantum key distribution over 90-km optical links installed in a field environment of Tokyo metropolitan area. IEEE J. Lightwave tech. 32, 141-151 (2013).

[40] Hirano, T., Yamanaka, H., Ashikaga, M., Konishi, T. and Namiki, R. Quantum cryptography using pulsed homodyne detection. Phys. Rev. A68, 042331 (2003).

[41] http://www.sequrenet.com/datasheets/datasheet_cygnus.pdf. Date of access: 01/03/2016.

[42] Shamir, A. How to share a secret. Communications of the ACM, 22, 612-613 (1979).

[43] Araki, T., Furukawa, J., Lindell, Y., Kazuma, A. N., Ohara, K., "High-Throughput Semi-Honest Secure Three-Party Computation with an Honest Majority", ACM CCS2016.

[44] https://jp.illumina.com/products/by-type/informatics-products/dragen-bio-it-platform.html Date of access: 06/24/2022.

[45] Wegman, M. N., and. Carter, J. L, "New Hash Functions and Their Use in Authentication and Set Inequality," J. Comput. System Sci. 22, pp.265-279, (1981).

[46] Alazab, M., "Forensic identification and detection of hidden and obfuscated malware,'" M.S. thesis, School Sci., Inf. Technol. Eng., Univ. Ballarat, Ballarat, VIC, Australia, (2012).

[47] Fujiwara, M., Masahiro Takeoka, M., and Sasaki, M. "Encryption key generator, encryption key generation program, and encryption key generation/erasure method," Japanese Patent Application No. 2021-051694, (2021).